\documentclass[prd,showpacs,twocolumn,preprintnumbers,10pt,aps]{revtex4-1}

\usepackage{amsmath,amssymb,amsfonts,amsthm}
\usepackage{bm}
\usepackage{hyperref}
\usepackage{graphics}

\newcommand{\be}{\begin{equation}}
\newcommand{\ee}{\end{equation}}
\newcommand{\ba}{\begin{eqnarray}}
\newcommand{\ea}{\end{eqnarray}}
\def\bs{\begin{subequations}}
\def\es{\end{subequations}}
\def\a{\alpha}

\def\la{\lambda}

\def\e{\epsilon}

\def\om{\omega}

\def\vr{\varrho}

\def\cF{{\cal F}}

\def\cL{{\cal L}}

\def\cS{{\cal S}}
\def\cV{{\cal V}}
\def\Om{\Omega}
\def\p{\partial}

\newcommand{\Eq}[1]{(\ref{#1})}
\def\lp{\ell_{\rm Pl}}

\def\rmd{d}
\def\rmi{i}
\def\x{q}
\def\ds{d_{\rm S}}
\def\dh{d_{\rm H}}
\newcommand{\doin}[2]{\href{http://dx.doi.org/#1}{#2}}
\newcommand{\arX}[1]{\href{http://arxiv.org/abs/#1}{{\ttfamily arXiv:#1}}}

\begin{document}

\title{Discrete to continuum transition in multifractal spacetimes}
\author{Gianluca Calcagni}
\affiliation{Max Planck Institute for Gravitational Physics (Albert Einstein Institute)
Am M\"uhlenberg 1, D-14476 Golm, Germany}

\date{June 1, 2011}

\begin{abstract}
We outline a field theory on a multifractal spacetime. The measure in the action is characterized by a varying Hausdorff dimension and logarithmic oscillations governed by a fundamental physical length. A fine hierarchy of length scales identifies different regimes, from a microscopic structure with discrete symmetries to an effectively continuum spacetime. Thanks to general arguments from fractal geometry, this scenario explicitly realizes two indirect or conjectured features of most quantum gravity models: a change of effective spacetime dimensionality with the probed scale, and the transition from a fundamentally discrete quantum spacetime to the continuum. It also allows us to probe ultramicroscopic scales where spectral methods based on ordinary geometry typically fail. Consequences for noncommutative field theories are discussed.
\end{abstract}

\preprint{\doin{10.1103/PhysRevD.84.061501}{Phys.\ Rev.\ D {\bf 84}, 061501(R) (2011)} \qquad [\arX{1106.0295}]}
\preprint{AEI-2011-030}
\maketitle


The idea that spacetime is fractal has been enticing the community for almost 40 years. The reasons beyond this concept and the way to realize it greatly varied from author to author \cite{ear,Ey89}, but the common denominator was a general lack of contact with physical applications. This may be due to the too-formal or too-heuristic character of the models, or simply because the full consequences of a spacetime with fractal geometry were not fully appreciated. Recent developments in quantum gravity stressed that theories on effective spacetimes with a certain fractal-like feature are typically ultraviolet (UV) finite \cite{spe}; we mention causal dynamical triangulations (CDT), asymptotic safety, Ho\v{r}ava-Lifshitz gravity, and spin foams. This feature is \emph{dimensional flow}: the spacetime spectral dimension changes from $\ds\leq2$ at small scales to $\ds=4$ at large scales. The smaller effective dimension in the UV is responsible for the finiteness (or renormalizability) of the models. Although its physical consequences are clear by now, this phenomenon is an indirect property with no obvious origin. Moreover, the spectral dimension was calculated at different individual scales, but no control has been exercised over dimensional flow as a whole. In an attempt to realize it as a direct and controllable property of spacetime, in \cite{fr123} a class of field theories was constructed where the action displays an exotic measure with anomalous scaling. However, there were several unclear points, including the relation between this type of continuum measure and genuine fractal geometry, the way to realize dimensional flow concretely, and the role of symmetries. Starting from the same motivations of \cite{fr123}, here we take the lessons from fractal geometry further seriously, and construct a nontrivial geometry via fractional calculus. The latter is an extremely rich compromise between full-fledged fractal geometry (to which it is related by precise approximation schemes \cite{frc}) and phenomenological continuum scenarios. Not only does this model allow us to answer to most of the elementary questions raised in \cite{fr123}, but it is also endowed with a surprising wealth of novel physical (and calculable) features with, possibly, long-range consequences for quantum gravity and noncommutative field theories. First, dimensional flow is implemented in a genuine all-scale fashion by a multifractal, self-similar measure. Second, the dimension of spacetime in the infrared is four by geometric requirements, not by assumption. Third, in the UV the theory is equipped with discrete symmetries which progressively melt into a continuum structure as the scale increases; this transition is only conjectured in discrete quantum gravity approaches, and is crucial to make them viable models of Nature. Fourth, at ultramicroscopic scales we are still in full control of the geometry, while methods based on intuitive geometric properties were unable to enter this regime of ``fuzzy'' spacetime. At these scales, connections with CDT and noncommutative geometry begin to emerge. 

All these achievements simply stem (i) from an application of the standard lore of fractal geometry and fractional calculus to a field theory context, and (ii) from the consequent reinterpretation of these tools in the spacetime arena. Mathematical details of the model will be given in two longer publications \cite{fra5}. Presently, we sketch the global physical picture for the general reader, with the hope to elicit interest in its applications. Contrary to \cite{fr123}, we do not include gravity and concentrate on the fractional analogue of Euclidean and Minkowski spaces.

\emph{Fractal regime.} Proceeding from small to large scales, at a deep microscopic level spacetime is assumed to be a multifractal structure $\cF$ described by standard tools of fractal geometry \cite{fge1,fge2}. This set is defined by a countable (possibly finite) number of transformations $\cS_i$ which leave $\cF$ invariant. Fractals admit both an implicit definition or can be conceived as embedded in an ambient space. A prototypical example is given by self-similar sets. Let the embedding space be spanned by coordinates $x$, and let $\cS_i(x)=\la_i x+a_i$ be $N$ similarities, where $\la_i>0$. As it happens for fractals, these symmetries are discrete; a one-dimensional case is the middle-third Cantor set, which is invariant under scaling transformations of strictly-fixed ratios $\la_{1,2}=1/3$. $\cF$ is constructed taking sequences of similarities and the intersection of sets $\cF_k=\cS_{i_1}\circ\cdots\circ \cS_{i_k}(U)$ for any $U\supset \cF$. Let $0<g_i<1$ be $N$ probabilities such that $\sum_i g_i=1$. One can imagine to distribute a mass on sets $\cF_k\subseteq\cF$ by dividing it repeatedly in $N$ subsets of $\cF_k$, in the ratios $g_1\,:\cdots:\,g_N$. This defines a self-similar measure $\vr$ with support $\cF$, such that $\vr(\cF_k)=g_{i_1}\dots g_{i_k}$ and, for all sets $A\subseteq\cF$,
\be\label{ssm}
\vr(A)=\sum_{i=1}^N g_i\,\vr[\cS_i^{-1}(A)]\,.
\ee
There is a highly nontrivial interplay between points ``inside'' the fractal and its boundary; this topological information is encoded in the so-called harmonic structure, and it is instrumental for a rigorous definition of the Laplacian operator on $\cF$. Together, geometric (i.e., symmetry) and harmonic structures determine the Hausdorff dimension of the set at any given scale, plus other parameters later to appear. If spacetime is multifractal, the support of the measure may be highly disconnected. We identify this fundamental regime by a generic self-similar measure \Eq{ssm}, without specifying its details. A field theory on such a set is formally described by the action
\be\label{genac}
S=\int \rmd \vr(x)\,\cL\,,
\ee
where the Lagrangian density $\cL$ contains any suitable choice of fields and differential operators allowed by the symmetries. Indeed, one can build a field theory on a fractal \cite{Ey89}, but there are two issues with that. On one hand, the mathematical status of any such construction has not been completely developed yet and, on the other hand, it is technically hard to make contact with manageable physics. This is the point where fractional calculus comes into play.

\emph{Log oscillations.} A fundamental result of spectral theory on deterministic fractals is that the heat kernel on $\cF$ displays logarithmic oscillations of period $\la_\om$ \cite{fge2}. Intuitively, this is due to the high degree of symmetry of $\cF$, where symmetry parameters and probability weights are fixed by default. These symmetries are unknown to continuous systems or artificially discrete systems such as lattices, and go under the name of \emph{discrete scale invariance} (DSI). A DSI is a dilation transformation under arbitrary powers $\la_\om^n$ of a preferred, special scaling ratio $\la_\om$ \cite{Sor98}. Log-periodicity and the associated DSI appear, e.g., in analyses of earthquakes and financial crashes, out-of-equilibrium and quenched disordered systems, phase transitions, and L\'evy flights. Such systems do have a fractal structure, but the associated spectral functions have an oscillatory behaviour, unusual from the point of view of conventional geometry. However, geometry is indeed well defined, and it is the purpose of this letter to implement these basic ideas in a model of spacetime. In the most general case, the number of the scaling ratios $\la_\om$ is infinite and governed by a multiplicity of frequency modes $\om$, determined by the geometric and harmonic structures. The periodic function in the heat kernel is not real-valued and this corresponds to an effective complex dimension of the fractal $\cF$ (on purpose, we do not specify which definition of dimension; see \cite{fra5}). To illustrate the main features of the proposal, it is sufficient to consider a measure $\vr_{\a,\om}(x)$ characterized by: (i) a set of nonnegative modes $\om\geq 0$, (ii) a real parameter $0\leq\a\leq1$, (iii) a combination of real coefficients $A_{\a,\om}$ and $B_{\a,\om}$ such that the measure is real-valued (then, we say the measure is self-conjugate). The form of these coefficients can be specified by fractional calculus and is not important here. For each direction $x=x^\mu$ in spacetime, at any given $\a$ and $\om$ we have
\bs\label{kcom2}\ba
\vr_{\a,\om}(x) &=& \frac{x^\a}{\Gamma(\a+1)}\left[1+A_{\a,\om}\cos\left(\om\ln\frac{x}{\ell_\infty}\right)\right.\nonumber\\
&&\left.+B_{\a,\om}\sin\left(\om\ln\frac{x}{\ell_\infty}\right)\right]\,,
\ea
where $\ell_\infty$ is a fundamental scale (or, in other words, a microscopic cutoff) forcefully introduced to make the arguments dimensionless. This measure is natural in fractional calculus of order $\a+\rmi\om$ and well approximates many features of fractal sets. While real-order measures realize either random fractals or deterministic fractals in the limit where most of the probability weight is concentrated in one portion of the set, complex-order calculus also accounts for the oscillatory structure \cite{frc}.

Given an embedding space with $D$ topological dimensions, the total integration measure reads
\be\label{osr}
\rmd\vr(x) = \sum_\a g_\a\sum_\om\prod_\mu \rmd\vr_{\a,\om}(x^\mu)\,,
\ee\es
where we also sum (or integrate) over $\a$. For each $\vr_{\a,\om}$, oscillations are governed by a dimensionless scale 
\be\label{scale0}
\la_\om=\exp(2\pi/\om)\,.
\ee
Notice the nonperturbative dependence on the frequency. The oscillatory part of Eq.~\Eq{kcom2} is log-periodic under the discrete scaling transformation $\ln(x/\ell_\infty)\to\ln(x/\ell_\infty)+2\pi n/\om$, $n=0,1,2,\dots$, 
 implying
\be\label{dsi}
x\,\to\, \la_\om^n x\,,\qquad n=0,1,2,\dots\,.
\ee
The characteristic (as opposed to fundamental) physical scale associated with $\la_\om$ is $\ell_\om=\la_\om\ell_\infty>\ell_\infty$. In the Euclidean case, $\mu=1,\dots,D$ and the embedding space is the positive orthant $\mathbb{R}^D_+$ with the origin, $x^\mu\geq 0$. In the case with Lorentzian signature, $\mu=0,\dots,D-1$ and the embedding space is the positive orthant of Minkowski spacetime. The sets $x^\mu=0$ and $x^\mu=+\infty$ are the boundary of the embedding space and of the set $\cF$. A fractional spacetime $\cF$ is defined by the embedding, the measure \Eq{kcom2}, a calculus which determines the differential operators in the Lagrangian density $\cL$, and a set of symmetries.

Fractional field theories can be regarded as effective, i.e., ``hydrodynamical'' approximations of microscopic discrete theories known to display dimensional flow, such as those of \cite{spe}. On the other hand, our model can propose itself also as fundamental, in which case one should take care of its UV finiteness.

In complex self-conjugate fractional models, there exists a hierarchy of length scales $\ell_\infty<\{\ell_\om\,:\,0\leq\om\leq \om_*\}<\ell_*$, where only $\ell_\infty$ is fundamental. The others will be called just ``characteristic'' and divide nonperturbatively different regimes. We describe qualitatively each of these regimes, leaving the details to \cite{fra5}. For simplicity, we consider only two modes, the zero-mode $\om=0$ and one nonvanishing positive mode $\om>0$.

\emph{Boundary-effect regime.} Here geometry is given by the continuum approximation \Eq{kcom2} with discrete scale invariance, but boundary effects are important: this happens when $x/\ell_\infty\sim 1$. Expanding Eq.~\Eq{kcom2} around this point and dropping a constant term, we have $\vr_{\a,\om}(x) =C_{\om,\a}\ln x+O(x/\ell_\infty)$, for some calculable, real, finite normalization constant $C_{\om,\a}$, so that the measure becomes
\be\label{noncor}
\vr(x)\sim \ln x\,,\qquad \ell\sim\ell_\infty\,.
\ee
This is not the same as taking the limit $\om\to\infty$, which is not well defined. Thus, the effective integration measure is of the form $\rmd\vr_{\a,\om}(x)\sim\ v_{\rm BE}(x)\,\rmd^D x:=(x^0x_1\cdots x_{D-1})^{-1}\rmd^D x$. This opens up the possibility to link together fractional and noncommutative spaces, in particular $\kappa$-Minkowski. Imposing a cyclicity-inducing measure weight $v_\kappa(x)$ in $\kappa$-Minkowski yields the conditions $\p_i[x^i v_\kappa(x)]=0$, $\p_0 v_\kappa(x)=0$. If one further imposes rotational symmetry, in $D-1$ dimensions, one obtains $v_\kappa(x)=|\mathbf{x}|^{1-D}$ \cite{AAAD}. However, this is not motivated by strict physical arguments, so another solution is $v_\kappa(x)=v_{\rm BE}(x)$ without the time coordinate. It is natural to establish a relation between $\kappa$-Minkowski and fractional models in the boundary-effect regime, with integer time direction. The fundamental scale of $\kappa$-Minkowski (what noncommutativists would call ``the Planck length'') is then identified with $\ell_\infty = \lp$. An independent argument leading to the same conclusion is given in \cite{fra5}. Since the period of the oscillations is given by the geometry of the set, we see the suggestive possibility to obtain the Planck length purely from symmetry.

\emph{Oscillatory transient regime.} In the range $\ell_\infty<\ell\ll\ell_*$, one should take the full form of Eq.~\Eq{osr}. The notion of dimension is defined only as an average over log-oscillations. Since the main symmetry of the theory is DSI, scenarios with self-conjugate measures can be regarded as nonperturbative, discrete models of spacetime.  Consistently with \Eq{dsi}, to get the continuum limit one should send the frequency to zero from above, so that the length cutoff vanishes: $\ell_\om\to 0$ as $\om\to 0^+$.

\emph{Multifractional regime.} At mesoscopic spacetime scales $\ell_\om\ll\ell\lesssim \ell_*$, one neglects boundary and oscillatory effects. The geometry is multifractional. To see this, one notices that at scales much larger than the period one can take the average of the measure $\vr_\a(x):= \langle\vr_{\a,\om}(x)\rangle\propto x^\a$. Then, the average of the oscillations is zero and one remains only with the zero mode. The total integration measure is
\be
\rmd\vr(x)\sim \sum_\a g_\a\rmd\vr_\a(x)\,,\qquad \ell_\infty\ll\ell\lesssim \ell_*\,.\label{mufr}
\ee
Picking only the zero mode corresponds to randomize the fractal structure. For a fixed $\a$, the Hausdorff dimension of spacetime is $\dh=D\a\leq D$. This means that the Euclidean volume of a $D$-ball of radius $R$ scales as $\cV\sim R^{D\a}$; this is apparent from the scaling of the measure, but it is found rigorously in \cite{fra5}. At this point, we have to quote some results obtained in that paper.

(i) Just like the oscillatory structure in the full measure, the sum over $\a$ in Eq.~\Eq{mufr} is motivated by fractal geometry. In particular, multifractal measures are of this form \cite{fge1}, where the $g_\a$ are nothing but the probability weights of Eq.~\Eq{ssm}, labeled by $\a$. From the perspective of field theory, the $g_\a$ are coupling constants attached to different operators, and the total multifractional action does coincide, quite naturally, to what one would get from renormalization group (RG) arguments: $S=\sum_\a g_\a \int\rmd\vr_\a(x)\,\cL_\a$, where $\cL_\a$ contains a finite set of operators. (ii) The range of $\a$ is dictated by the requirement that fractional spacetime has a natural norm. This happens if $\a\geq 1/2$. (iii) The dimension of spacetime changes with the scale. The coefficients $g_\a$ are dimensionful, in order for $S$ to be dimensionless. Then, they determine the scale at which geometry changes. Consider a simplified model with one such scale $\ell_*$. A $D$-ball of radius $R$ would have volume $\cV= \ell_*^{D\a_1}[\Om_{D,\a_1} (R/\ell_*)^{D\a_1}+\Om_{D,\a_2}(R/\ell_*)^{D\a_2}]$, where $\a_1<\a_2$ and $\Om_{D,\a}$ are dimensionless coefficients. For a small ball ($R\ll\ell_*$), $\cV\sim R^{D\a_1}$, while for a large ball ($R\gg\ell_*$), $\cV\sim \tilde R^{D\a_2}$, where $\tilde R=R \ell_*^{-1+\a_1/\a_2}$ is the radius measured in macroscopic length units. 

(iv) In this phase, continuous symmetries emerge. In fact, while the zero-mode part of the measure clearly breaks ordinary Poincar\'e invariance, it is invariant under nonlinear transformations of the embedding coordinates $x^\mu$ which preserve the fractional line element \cite{fra5}. As a matter of fact, fractional spacetimes with fixed $\a$ admit a very natural set of ``geometric coordinates'' given by (index $\mu$ omitted) $\x(x):=\vr_\a(x)=x^\a/\Gamma(\a+1)$. Since these coordinates coincide with the measure along each direction, the measure in the total action at each $\a$ is invariant under the affinity
\be\label{fpotra}
{\x'}^\mu(x) = \tilde\Lambda_\nu^\mu \x^\nu(x) +\tilde a^\mu\,,
\ee
where $\tilde a^\mu$ is a constant vector and $\tilde\Lambda_\nu^\mu\tilde\Lambda_\mu^\rho=\delta_\nu^\rho$ are Lorentz matrices. In this respect, fractional spacetimes can be argued to be self-affine sets in geometric coordinates. If the fractional symmetry is imposed also to $\cL_\a$, one obtains a ``fractional symmetry scenario'' where the proliferation of operators is severely restricted scale by scale. This situation is different from Lorentz-invariant field theory, which is endowed with a symmetry group fixed all along the RG flow. A hybrid scenario, which we call ``fractional/integer,'' maintains Eq.~\Eq{fpotra} only as a symmetry of the measure, while the Lagrangian $\cL_\a$ is made covariant under ordinary Poincar\'e transformations. This corresponds to a different definition of the UV Laplacian/d'Alembertian, which may not be unique \cite{fra5}.

(v) The UV and infrared Hausdorff dimensions of spacetime are tightly related to each other. One of the original motivations of \cite{fr123} was to obtain a theory where the UV dimension be 2, in accordance with the phenomenon of dimensional flow in independent quantum gravity scenarios \cite{spe}. The deep reasons why number 2 is special lie in the UV finiteness of all these models, but there also exists a heuristic argument singling it out of all the possibilities. Including Planck's constant $\hbar$, the electron charge $e$, Newton's constant $G$, and the speed of light $c$, one can construct a dimensionless constant in a spacetime of Hausdorff dimension $\dh$ as $C=\lp^{2(3-\dh)} e^{\dh-2} G^{(\dh/2)-1} c^{2(2-\dh)}$, where $\lp=\sqrt{\hbar G/c^3}$ is the Planck length. Remarkably, in $\dh=2$ the fundamental constant coincides with (the square of) the Planck length, $C=\lp^2$, while all the other couplings disappear. Now, it turns out that, at least for the fractional/integer symmetry scenario, the theory has a critical point at $\a=\a_*=2/D$, corresponding to $\dh=2$. If $\a_*$ is also the lowest possible $\a$, where dimensional flow stops, then one must have $D=4$. Thus, four dimensions are selected by geometry arguments. To summarize, at microscopic scales much larger than a log-period $\ell_\om$ but smaller than $\sim\ell_*$, spacetime is effectively two-dimensional with well-defined fractional geometry, given by the measure
\be\label{2dme}
\vr(x)\sim \vr_{1/2}(x)\propto x^{1/2}\,,\qquad \ell_\om\ll\ell\lesssim\ell_*\,.
\ee

\emph{Classical regime.} Finally, at scales larger than the characteristic scale $\ell_*$, ordinary field theory on Euclidean/Minkowski spacetime is recovered:
\be
\vr(x)\sim \vr_{1}(x)=x\,, \qquad \ell\gg\ell_*\,.
\ee
The number of topological dimensions is theoretically constrained to be four. In this regime, the theory is almost Poincar\'e invariant in the standard sense, the Hausdorff dimension of spacetime is $\dh=4-\e$, and Euclidean geometry in local inertial frames gets tiny corrections. Bounds on the parameter $\e$ can be taken from the literature of dimensional regularization models; one roughly obtains $|\e|<10^{-8}$ at scales $\ell\sim 10^{-15}\,{\rm m}$. The characteristic scale $\ell_*<10^{-18}\,\mbox{m}$ can be constrained by particle physics observations.

\emph{Discussion.} Thanks to a close inspection of the fractal properties of fractional models, we have been able to formulate a concrete realization of dimensional flow. This phenomenon was only conjectured in previous papers on fractal spacetimes, due to the difficulty in understanding dimensional properties even at a fixed scale (here, fixed $\a$). The results obtained here should allow us to begin a detailed study of the multifractional regime with RG techniques. Apart from the natural realization of a discrete-to-continuum transition in spacetime, other challenging issues in theoretical physics could be reinterpreted under a novel perspective. First, due to the nonperturbative exponential law \Eq{scale0}, it may be possible to relax the hierarchy problem. Second, the extension of fractal models to gravity are likely to have notable consequences on inflation (which can be replaced by the alternative mechanism of dimensional flow) and the big bang problem (what is the role of measure oscillations near the singularity?). Third, there exists a mapping between general fractional models with exotic power-law measure $\vr_\a$ and noncommutative spacetimes with algebras more general than $\kappa$-Poincar\'e \cite{ACOS}. Finally, oscillatory measures can have concrete applications in quantum gravity approaches such as CDT. One might be able to provide further geometrical insights into the crumpled phase (phase B) and the branched-polymeric phase (phase A) of the CDT phase diagram \cite{AGJJL}, corresponding, respectively, to configurations in the near-boundary regime (formally similar to the structureless limit $\a\to 0$ in the averaged measure) and to the oscillatory regime.


\end{document}